\documentclass{article}
\usepackage{INTERSPEECH2022}
\usepackage{amsmath,graphicx}

\usepackage{algorithm}
\usepackage{algpseudocode}
\usepackage{pifont}
\usepackage{amssymb}
\usepackage{bm}
\usepackage{multirow}
\usepackage{adjustbox}
\usepackage{tabularx}
\usepackage{tikz}
\usepackage{caption}
\usepackage{subcaption}
\usepackage{verbatim}
\usepackage{float}
\usepackage{commath}
\usepackage{url}
\usepackage{color, colortbl}

\definecolor{LightCyan}{rgb}{0.88,1,1}

\title{Time-domain Ad-hoc Array Speech Enhancement Using a Triple-path Network}
\name{Ashutosh Pandey$^{\textup{1}*}$, Buye Xu$^\textup{1}$, Anurag Kumar$^\textup{1}$, Jacob Donley$^\textup{1}$, Paul Calamia$^\textup{1}$ and DeLiang Wang$^\textup{2}$\thanks{*Work done during internship at Reality Labs Research at Meta.}}
\address{$^\textup{1}$Reality Labs Research at Meta, Redmond, WA USA\\
$^\textup{2}$Department of Computer Science and Engineering, The Ohio State University, USA}
\email{}
\begin{document}
\ninept
\maketitle
\begin{abstract}
Deep neural networks (DNNs) are very effective for multichannel speech enhancement with fixed array geometries. However, it is not trivial to use DNNs for ad-hoc arrays with unknown order and placement of microphones. We propose a novel triple-path network for ad-hoc array processing in the time domain. The key idea in the network design is to divide the overall processing into spatial processing and temporal processing and use self-attention for spatial processing. Using self-attention for spatial processing makes the network invariant to the order and the number of microphones. The temporal processing is done independently for all channels using a recently proposed dual-path attentive recurrent network. The proposed network is a multiple-input multiple-output architecture that can simultaneously enhance signals at all microphones. Experimental results demonstrate the excellent performance of the proposed approach. Further, we present analysis to demonstrate the effectiveness of the proposed network in utilizing multichannel information even from microphones at far locations.

\end{abstract}
\section{Introduction}
\label{sec:intro}
Multi-channel speech enhancement is concerned with improving the intelligibility and quality of noisy speech by utilizing signals from microphone arrays. Traditional approaches use linear spatial filters in a filter-and-sum process designed to preserve signals of the target source \emph{(e.g.}, constrained to be undistorted) and attenuate signals from interference (e.g. minimize noise variance), the latter of which are often separated from the target source in the spatial domain~\cite{gannot2017consolidated}. Some of these approaches leverage spatial correlations of speech and noise to determine filter coefficients, and hence, are convenient to use with unknown array geometries~\cite{markovichgolan2015optimal}.

Supervised speech enhancement using deep neural networks has achieved remarkable success and popularity in the last few years~\cite{wang2017supervised}. On the multi-channel front, DNNs have been extensively studied with fixed array geometries~\cite{erdogan2016improved, heymann2016neural, wang2018combining, wang2018multi, wang2020multi, tolooshams2020channel, gu2019end, tzirakis2021multi, pandey2022tparn}. However, neural-network based speech enhancement with ad-hoc arrays, where microphone geometries and distributions might not be known, has not received much attention and remains little explored. Ad-hoc array processing offers considerable flexibility compared to microphone arrays with fixed geometries and can play a crucial role in enabling audio and speech applications in the real world. For example, it is amenable to larger apertures in the context of wearables as they are not restricted to the size of small wearable devices. Moreover, methods developed for ad-hoc array processing can be easier to use, adapt, and transfer in different situations, as by design these methods are expected to work in situations where microphone numbers and distributions might not be known. 


However, ad-hoc array processing using deep neural networks remains a challenging problem. First, it requires a network to be able to process a multi-channel signal with an unknown number of microphones at random locations and in any order. In other words, the network should be invariant to the number, geometry and the order of microphones. Second, different microphones can be asynchronous. A systematic approach to solve the first problem is designing networks with processing blocks that are number and permutation invariant, such as global pooling and self-attention~\cite{zaheer2017deep}. The second problem can be solved by using correlation-based approaches ~\cite{liu2008sound, araki2018meeting, yoshioka2019meeting}. 

Some recent works have investigated DNNs for ad-hoc array processing~\cite{luo2020end, wang2020neural, tzirakis2021multi, furnon2021dnn, furnon2021attention, zhang2021deep}. Luo et al.~\cite{luo2020end} proposed a novel transform-average-concatenate module to deal with unknown number and order, and graph neural networks were investigated for distributed arrays by Tzirakis et al.~\cite{tzirakis2021multi}. Wang et al.~\cite{wang2021continuous} proposed a spatio-temporal network where a recurrent network was used for temporal modeling and self-attention was used for spatial modeling. The output corresponding to the reference microphone was obtained by using a global pooling layer in the end. Deep ad-hoc beamforming proposed in~\cite{zhang2021deep} utilizes a two-stage approach: first select the top $k$ microphones and then use the selected $k$ signals for $k$-microphone speech enhancement in the second stage. Furnon \emph{et al.} ~\cite{furnon2021dnn} proposed a DNN-based two-stage approach to distributed multi-channel Weiner filtering~\cite{bertrand2010distributed}, where a single node DNN is used for node-specific enhancement followed by a multi-node DNN for global enhancement.  

In this paper, we propose a triple-path network, \textit{TADRN: Triple-Attentive Dual-Recurrent Network}, for ad-hoc array processing in the time domain. The key idea in the TADRN design is to divide the overall processing into spatial processing and temporal processing and use self-attention for spatial processing. Using self-attention for spatial processing makes the network invariant to the order and the number of microphones. The spatial processing is followed by a dual-path attentive recurrent network (ARN), a recurrent network augmented with self-attention~\cite{pandey2020dual, pandey2022self}, for temporal processing. The temporal modeling is performed independently for all channels, by first dividing a signal into smaller chunks and then using separate ARNs to process intra-chunks and inter-chunks data. The intra-chunk processing enables local learning whereas inter-chunk processing helps capture global dependencies. Thus, the TADRN becomes a triple-path attention framework: operating on channels (inter-channel), within chunks of audio (intra-chunk) and across the chunks (inter-chunks). Moreover, the intra-chunk and inter-chunks learning are aided by recurrent architectures. 


The TADRN design is similar to a recently proposed triple-path attentive recurrent network (TPARN) for fixed array processing ~\cite{pandey2022tparn}. However, there are two key differences. First, TADRN uses self-attention across channels whereas TPARN uses ARN. While theoretically, TPARN and TADRN are both capable of handling an unknown number of microphones, the use of self-attention across channels makes TADRN order-invariant, and hence, more suitable for ad-hoc array processing. Second, the processing order of underlying blocks, intra-chunk, inter-chunk, and inter-channel blocks, is different in TADRN, determined based on empirical observations.

TADRN is a multiple-input and multiple-output (MIMO) architecture that can simultaneously enhance signals from all microphones. Our empirical evaluations shows that the proposed TADRN approach can outperform prior methods for speech enhancement using ad-hoc arrays. Moreover, we analyze its behavior and attempt to provide important insights into the method. More specifically, we show that TADRN can improve enhancement by leveraging additional randomly placed microphones, even at locations far from the source. Additionally, large improvements in objective scores are observed when poorly placed microphones (far from source) in the scene are complemented with  more effective microphone positions, such as those closer to a target source.

\begin{figure}[!t]
\centering
\includegraphics[width=\columnwidth, keepaspectratio]{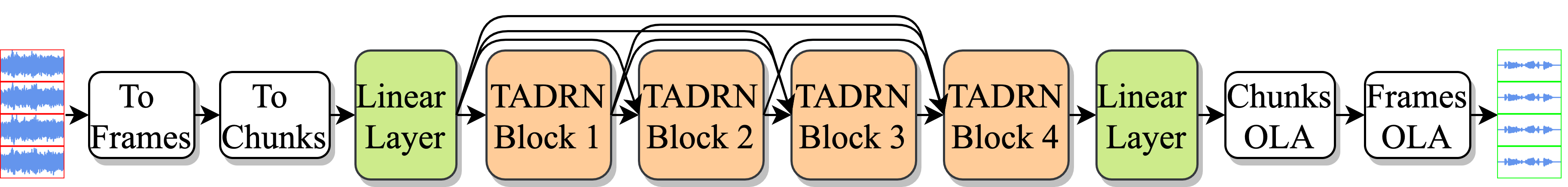}
\caption{\it The proposed TADRN architecture for ad-hoc array multichannel speech enhancement.}
\label{fig:tadrn}
\end{figure}
\section{Problem Definition}
A multi-channel noisy signal \scalebox{0.98}{$\bm{X} = [\bm{x}_{1}, \dots, \bm{x}_{P}]\in \mathbb{R}^{P \times N}$} with $N$ samples and $P$ microphones is modeled as
\begin{equation}
\begin{adjustbox}{width=0.91\columnwidth}
$
\begin{split}
x_{p}(n) &= y_{p}(n) + z_{p}(n) \\
               &= h_{p}(n) \ast s(n) + z_{p}(n) \\
               &= g_{p}(n) \ast s(n) + [(h_{p} - g_{p})(n) \ast s(n) + z_{p}(n)] \\
               & = d_{p}(n) + u_{p}(n)
\end{split} 
$
\end{adjustbox}
\end{equation}
where $p = 1, 2 \dots P$ and $n = 0, 1, \dots N-1$. $\bm{y}_{p}$ and $\bm{z}_{p}$ respectively represent the reverberant speech and noise received at $p^{th}$ microphone, and $\bm{s}$ is the target speech at the sound source. $\bm{h}_{p}$ is the room-impulse-response (RIR) from the target source to the $p^{th}$ microphone, and $\bm{g}_{p}$ is the direct-path impulse response accounting for the free-field propagation of the sound. $\bm{d_{p}}$ is the direct-path signal from the speech source, and $\bm{u}_{p}$ denotes the overall interference in this paper, which includes the background noise and the reverberation of the target speech. The goal is to get a close estimate, $\bm{\hat{d}}_{r}$, of the direct-path signal in the predefined reference channel $r$, $\bm{d}_{r}$. 

For a fixed array case, the number of microphones ($P$) and the array geometry (\emph{e.g.}, a circular array with a fixed radius), is known beforehand and remains unchanged. However, for an ad-hoc array, the microphones can be randomly distributed in the environment. Neither the number nor the relative locations of the microphones can be assumed to be known \emph{a priori}. 

\begin{figure}[!t]
\centering
\includegraphics[width=\columnwidth, keepaspectratio]{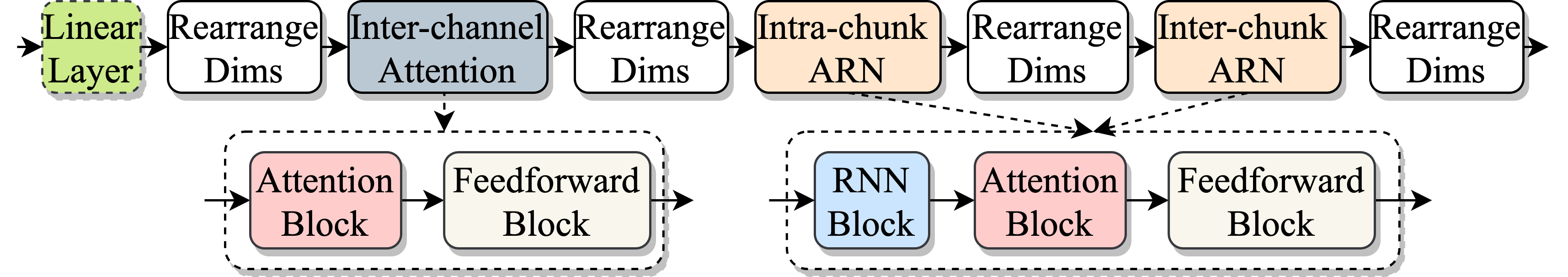}
\caption{Architecture of TADRN Block.}
\label{fig:tadrn_block}
\end{figure}

\section{Triple-attentive Dual-recurrent Network}
The full block diagram of TADRN is shown in Fig. \ref{fig:tadrn}. First, the multi-channel input signal $\bm{X}$ is converted into sequential frames using a frame size of $L$ samples and a frame shift of $K$ samples, leading to 3D tensor representation of the signal, $\bm{\text{T}}= [\bm{X}_{1}, \dots, \bm{X}_{T}] \in \mathbb{R}^{P \times T \times L}$, where $T$ is the number of frames. The consecutive frames are further grouped into chunks with a chunk size of $R$ and chunk shift of $S$ forming a 4D tensor $\mathcal{T} = [\bm{\text{T}}_{1}, \dots, \bm{\text{T}}_{C}] \in \mathbb{R}^{P \times C \times R \times L}$, where $C$ is the number of chunks.

The audio frames are first encoded to D-dimensional representations  using a linear layer, leading to $\bm{\text{E}} \in \mathbb{R}^{P \times C \times R \times D}$  as outputs from the linear layer. $\bm{\text{E}}$ is then processed using a stack of four TADRN blocks. Let $\bm{\text{B}}_{i}^{inp}$ and $\bm{\text{B}}_{i}^{out}$  denote the input and output of the $i^{th}$ block, respectively.  $\bm{\text{B}}_{1}^{inp} = \bm{\text{E}}$, $\bm{\text{B}}_{i}^{out} \in \mathbb{R}^{P \times C \times R \times D}$, and \scalebox{0.96}{$\bm{\text{B}}_{i}^{inp} = [\bm{\text{E}}, \bm{\text{B}}_{1}^{out}, \dots, \bm{\text{B}}_{i-1}^{out}] \in \mathbb{R}^{P \times C \times R \times i \cdot D}$} for $i > 1$.

The complete architecture of the TADRN block is shown in Fig. \ref{fig:tadrn_block}. It consists of an optional linear layer followed by inter-channel attention, intra-chunk ARN, and inter-chunk ARN. The linear layer is used for $i > 1$ to project a feature of size $i \cdot D$ to size $D$. The inter-channel attention comprises an attention block and a feedforward block, and the intera-chunk and inter-chunk ARNs comprise an RNN block, an attention block, and a feedforward block. 

The inputs to the inter-channel attention are first rearranged to tensors of shape $C \cdot R \times P \times D$.  These are then processed by treating the first, second, and third dimension respectively as the batch, sequence, and feature dimension. As a result, attention is applied across channels for spatial modeling. Similarly, the inputs to the intra-chunk ARN are reshaped to $P \cdot C \times R \times D$ to treat frames within a chunk as a sequence for the local temporal modeling, and  the inputs to the inter-chunk ARN are reshaped to shape $P \cdot R \times C \times D$ to treat different chunks as a sequence for learning global temporal characteristics. Finally, the output from the inter-chunk ARN is rearranged to the original shape of $P \times C \times R \times D$.

\begin{figure}[!t]
\centering
\includegraphics[width=0.65\columnwidth, keepaspectratio]{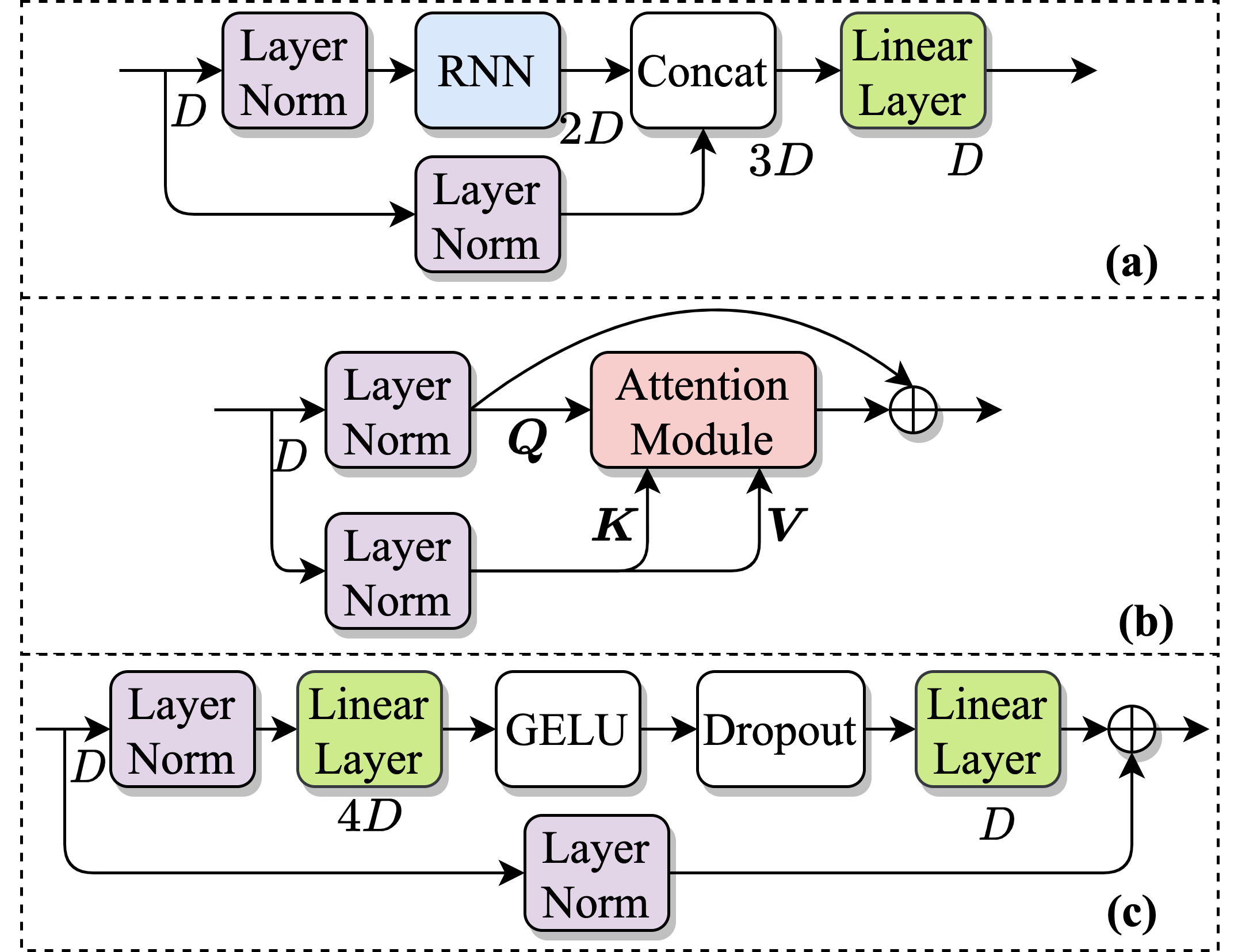}
\caption{\it (a) RNN block,  (b) Attention block, (c) Feedforward block.}
\label{fig:arn}
\end{figure}

The structure of the RNN block, the attention block, and the feedforward block in ARN is shown in Fig. \ref{fig:arn}. The inputs to all blocks are first split into two streams using two separate layer normalizations. The first stream in the attention block is used as query ($\bm{Q}$) and the second stream is used as key ($\bm{K}$) and value ($\bm{V}$) for the attention module. The outputs of the attention module are added to $\bm{Q}$ to form a residual connection. Additional details of the attention module can be be found in \cite{pandey2022tparn}. 

The first stream in the RNN block is processed using a RNN with a hidden size of $2D$. It is then concatenated with the second stream, and then projected to a size of $D$ using a linear layer. The first stream of the feedforward block is projected to a size of $4D$ using a linear layer with the GELU nonlinearity and dropout, projected again to the size of $D$ using another linear layer, and then added to the second stream to form a residual connection. 

\section{Experiments}
\subsection{Datasets}

We create an ad-hoc array dataset using speech and noises from the DNS challenge 2020 corpus\footnote{\footnotesize{\url{https://github.com/microsoft/DNS-Challenge/blob/master/LICENSE}}}~\cite{reddy2020interspeech}. We select speakers with one chapter and randomly split  90\% of speakers for training, 5\% for validation, and 5\% for evaluation. After this, for each utterance a random chunk of a randomly sampled length with an activity threshold (from script in~\cite{reddy2020interspeech}) greater than 0.6 is extracted. The length of utterances are sampled from [3, 6] seconds for training and [3, 10] seconds for test and validation. This results in a total of 53k utterances for training, 2.6k for validation, and 3.3k for test. Next, all the noises from the DNS corpus are randomly divided into training, validation and test noises in a proportion similar to that used for speech utterances.

The algorithm to generate spatialized multichannel ad-hoc array data from DNS speech and noises is given in Algorithm 1. We sample a room size and then sample $6$ locations inside the room for microphones, one location for the speech source and $5-10$ locations for noises. All the locations are sampled at least 0.5 m away from walls. We simulate the room-impulse-responses (RIRs) from each source location to all the microphone locations, and then convolve them with the speech and noise signals. Finally, the convolved speech and noises are added together using a random SNR value to create ad-hoc array multichannel noisy data. The SNR is calculated using total speech and noise energy at all microphones. In this case, the use of random locations for microphones is creating an ad-hoc array scenario. We use Pyroomacoustics~\cite{scheibler2018pyroomacoustics} which uses a hybrid approach where the image method with order 6 is used to model early reflections and ray-tracing is used to model the late reverberation. 
\setlength{\textfloatsep}{0.4cm}
\begin{algorithm}[!t]
\caption{Ad-hoc array dataset spatialization process.}
\begin{algorithmic}
\footnotesize
\For {\emph{split} in \{train, test, validation \}}
\For {speech utterances in \emph{split} }
\begin{itemize}
\item Draw room length and width from [5,10] m, and height from [3, 4] m;
\item Draw 6 microphone locations inside the room
\item Draw 1 speech source location inside the room;
\item Draw $N_{ns}$, number of noise sources, from [5, 10]
\item Draw $N_{ns}$ noise locations inside  the room
\item Sample T60 uniformly from [0.2, 1.3] seconds
\item Generate RIRs corresponding to speech source location and $N_{ns}$ noise locations for all microphone locations
\item Draw $N_{ns}$ noise utterances from noises in \emph{split} 
\item Propagate speech and noise signals to all mics by convolving with corresponding RIRs
\item Draw a value $snr$ from [-10, 10] dB, and add speech and noises at each mic using a scale so that the averaged SNR across all the microphone locations is $snr$;
\end{itemize}
\EndFor
\EndFor
\end{algorithmic}
\end{algorithm} 
\setlength{\floatsep}{0.4cm}

\subsection{Experimental settings}
All the utterances are resampled to 16 kHz.  We break the utterances into frames using $L=16$ and $K=8$. We use $R=126$ and $S=63$ to group frames into chunks. TADRN uses $4$ blocks with $D=128$ inside ARN. For the RNN in the ARN, we use bidirectional long short-term memory networks (BLSTMs) with the hidden size of $D$ in each direction. The dropout rate in the feedforward blocks of ARN is set to $5\%$. A phase constrained magnitude (PCM) loss over all channels is used for training ~\cite{pandey2022tparn, pandey2021dense}. All the models are trained for $100$ epochs on $4$-second long utterances with a batch size of $8$. For utterances longer than $4$ seconds, we dynamically extract a random chunk of 4 seconds during training. 

The automatic mixed precision training is utilized for efficient training~\cite{micikevicius2018mixed}. The learning rate is initialized with 0.0004 and is dynamically scaled to half if the best validation score does not improve in five consecutive epochs. The model with the best validation score is used for evaluation.

All of the models are trained on microphone arrays with $2$, $4$ , or $6$ channels, and evaluated on arrays with $1-6$ channels, where $1$, $3$, and $5$ are untrained numbers of microphones. During training, we first randomly sample $p$, the number of microphones from $\{2, 4, 6\}$, and then create a batch of training examples with $p$ microphones. This is done to avoid redundant computations.  

We compare TADRN with a recently proposed fiter-and-sum network with transform average and concatenate module (FasNet-TAC) for ad-hoc array processing~\cite{luo2020end}.  We also compare TADRN with three fixed-array baseline models: dense convolutional recurrent network (DCRN)~\cite{wang2020multi}, FasNet-TAC ~\cite{luo2020end}, and channel-attention Dense UNet (CA-DUNet)~\cite{tolooshams2020channel}.

The models are compared using three enhancement objective metrics: short-time objective intelligibility (STOI)~\cite{taal2011algorithm}, perceptual evaluation of speech quality (PESQ)~\cite{rix2001perceptual}, and scale-invariant signal-to-distortion ratio (SI-SDR) for signals from the first microphone channel. Note that the numbering of the microphone channels are randomly assigned. STOI is reported in percentage. 


\begin{table}[!h]
\centering
\caption{TADRN comparison with an ad-hoc array baseline model.}
\begin{adjustbox}{width=0.84\columnwidth}
\begin{tabular}{|c||c||c|c|c|c|c|c|c|}
\cline{3-9}
\multicolumn{1}{c}{} & \multicolumn{1}{c||}{} & Mix. & 1ch & 2ch & 3ch & 4ch & 5ch & 6ch \\
\hline
\hline
\multirow{2}{*}{ SI-SDR } & FASNET-TAC~\cite{luo2020end} & \multirow{2}{*}{ -12.2 } & -6.1 & -3.0 & -1.5 & -0.4 & 0.4 & 1.0 \\
& TADRN & & \textbf{-0.8} & \textbf{3.0} & \textbf{4.5} & \textbf{5.4} & \textbf{6.0} & \textbf{6.4} \\
\hline
\hline
\multirow{2}{*}{ STOI } & FASNET-TAC~\cite{luo2020end} & \multirow{2}{*}{ 60.8 } & 78.0 & 81.0 & 82.4 & 83.3  & 84.0 & 84.5 \\
& TADRN & & \textbf{83.3} & \textbf{87.3} & \textbf{88.9} & \textbf{89.9} & \textbf{90.5} & \textbf{90.9} \\
\hline
\hline
\multirow{2}{*}{ PESQ } & FASNET-TAC~\cite{luo2020end} & \multirow{2}{*}{ 1.40 } & 1.81 & 1.94 & 2.02 & 2.07 & 2.11 & 2.14 \\
& TADRN & & \textbf{2.12} & \textbf{2.38} & \textbf{2.51} & \textbf{2.58} & \textbf{2.64} & \textbf{2.68} \\
\hline
\end{tabular}
\end{adjustbox}
\end{table}

\begin{figure}[!t]
\centering
\includegraphics[width=\columnwidth, keepaspectratio]{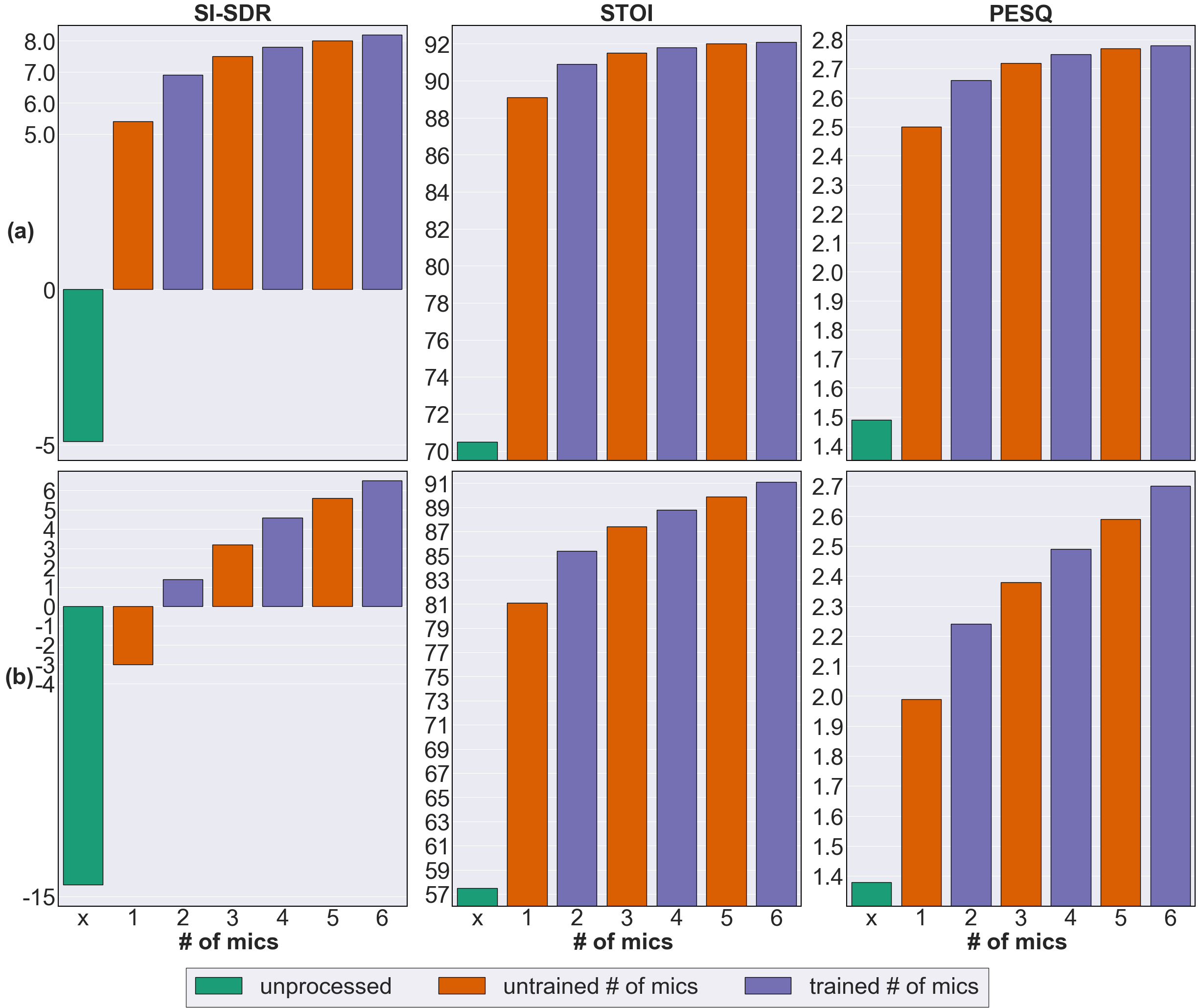}
\caption{\it TADRN performance with different number of microphones. a) Microphones sorted by increasing distance from the source, b) Microphones sorted by decreasing distance from the source.}
\label{fig:rnnt}
\end{figure}

\begin{figure}[!t]
\centering
\includegraphics[width=\columnwidth, keepaspectratio]{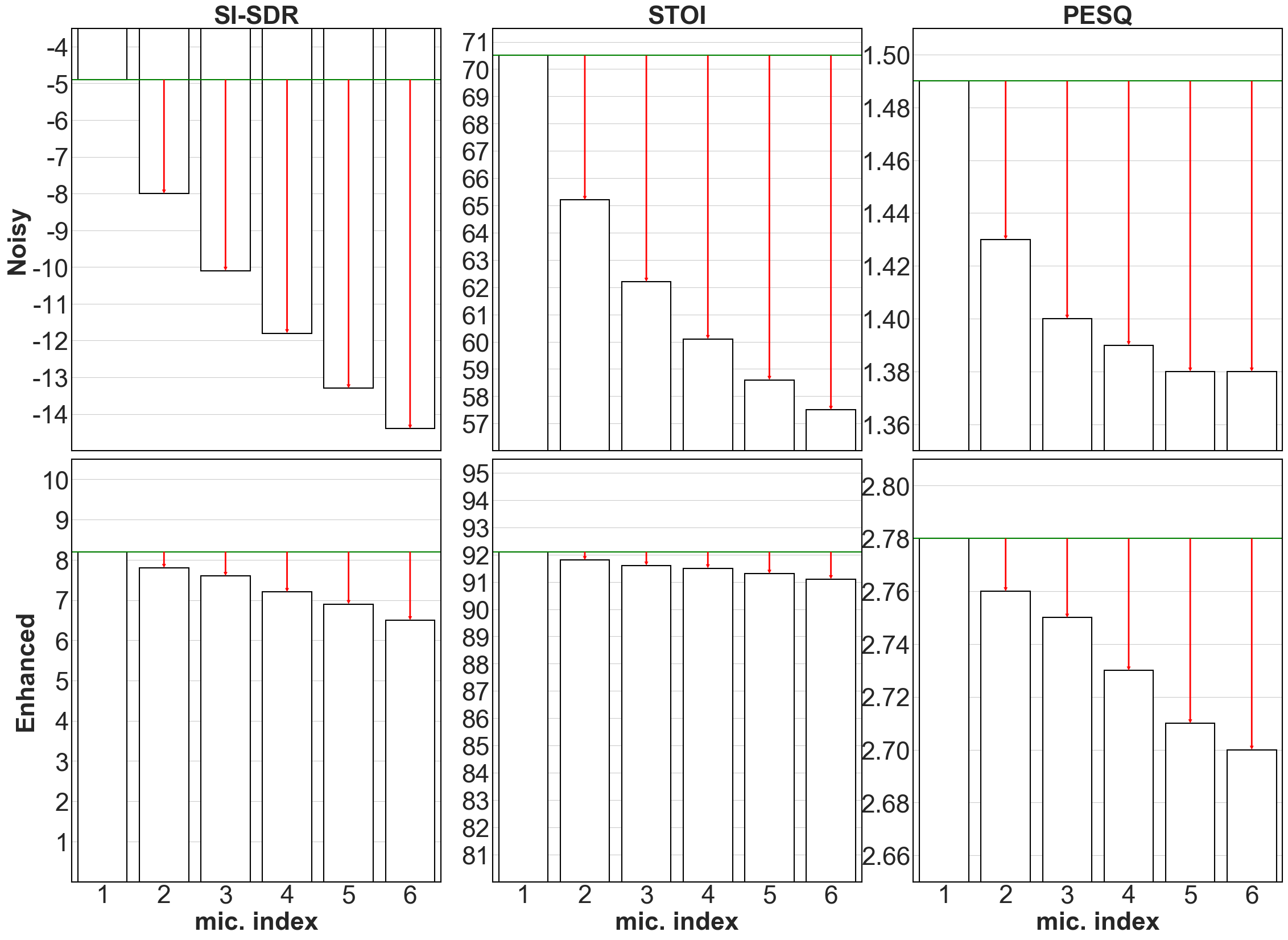}
\caption{\it An illustration of the gap reduction in the objective scores at different microphones after MIMO processing of TADRN. }
\label{fig:rnnt}
\end{figure}

\subsection{Experimental results}

\subsubsection{Comparison with Prior Works}
First, we compare TADRN with the baseline FasNet-TAC. Results with different numbers of microphones are given in Table 1. We observe that the performance of both the models improves gradually as the number of microphones are increased. However, TADRN results are consistently and significantly better than FasNet-TAC for all the cases. For instance, for the 6-channel case, TADRN outperforms FasNet-TAC by $5.4$ dB in  SI-SDR, $6.4\%$ in STOI, and $0.54$ in PESQ. Moreover, single-channel TADRN is able to outperform 3-channel FasNet-TAC. 

\subsubsection{Analysis of Impact of Microphone Locations}
Results in Table 1 indicate that TADRN can improve performance by adding more microphones at random locations inside a room. Because the microphones in an ad-hoc array can be well separated in space, the SNR at each microphone location could be significantly different. For example, SNR at locations closer to the speech source is generally higher than locations further away because of the decay of the sound energy. In order to understand how the local SNR of each microphone affects the array performance, we conduct two experiments. First, we sort microphones in the increasing order of distance from the source to simulate decreasing order of SNR (on average), and evaluate the signal improvement in the channel of the microphone closest to the source. This is visualized in Fig. 4 (a). We observe that TADRN gradually improves performance when more microphones are added at further distance, although the performance tends to saturate after four to five microphones. Second, in Fig. 4 (b), we analyze TADRN's behavior for a reverse case where arrays are sorted in the order of decreasing distance from the source. For this case, TADRN improves performance significantly and consistently with increasing number of microphones, as newly added microphones are closer to the source.

These results suggest that while TADRN naturally relies more on high SNR microphone channels, it can also leverage all other lower SNR channels to further improve the performance. 

\subsubsection{Analysis of MIMO}
TADRN is a MIMO architecture that can enhance signals from all microphones simultaneously. We analyze performance improvements for all microphones in an array sorted in the order of increasing distance from the source. This is visualized in Fig. 5, and an interesting behavior can be observed in the plot. The objective scores for unprocessed signals differ a lot at different microphones in an array. However, for enhanced signals, the difference between objective scores becomes very small. For instance, for the unprocessed case, the difference between the $1^{st}$ and the $6^{th}$ microphone is close to $9$ dB for SI-SDR and $13\%$ for STOI. For the enhanced case, however, the difference is less than $2$ dB for SI-SDR and  close to $1\%$ for STOI. We also note that the difference in PESQ is decreased by a relatively smaller amount. We think that this may be because there is only a small difference in PESQ for the unprocessed case. This analysis illustrates that in the proposed MIMO learning framework, poorly placed microphones at far locations from the source are benefitting the most by utilizing better signals from microphones at closer locations.   

\subsubsection{Application to Fixed-Geometry Arrays}
Finally, we evaluate a TADRN trained on ad-hoc arrays for a 4-channel circular array dataset from the DNS corpus~\cite{pandey2022tparn}. Also, we compare with existing models trained specifically for a 4-channel circular array. Objective scores are given in Table 2. We can see that even though TADRN is far from TPARN, it is able to outperform two fixed-array baselines, CA-DUNet and FasNet-TAC. It is better than DCRN in SI-SDR but worse in STOI and PESQ.    

\begin{table}[!t]
\centering
\caption{TADRN Performance on fixed-array dataset}
\begin{adjustbox}{width=0.64\columnwidth}
\begin{tabular}{|c|c|c|c|c|}
\hline
& Test Metric & SI-SDR & STOI & PESQ \\
\hline
Type & Unprocessed & -7.6 & 63.8 & 1.38 \\
\hline
\hline
fixed & CA-DUNet~\cite{tolooshams2020channel} & 2.7 & 82.1 & 1.93 \\
fixed & DCRN~\cite{wang2020multi} & 4.6 & 90.1 & 2.57 \\
fixed & FasNet-TAC~\cite{luo2020end} & 4.7 & 86.5 & 2.26 \\
\hline
\hline
fixed & TPARN~\cite{pandey2022tparn}& \textbf{8.4} & \textbf{91.9} & \textbf{2.75} \\
ad-hoc & TADRN & 5.6 & 88.1 & 2.41 \\
\hline
\end{tabular}
\end{adjustbox}
\end{table}

\section{Conclusions}
We have proposed a  triple-attentive dual-recurrent network for ad-hoc array multichannel speech enhancement in the time-domain. TADRN is designed by extending a single-channel dual-path model to a multichannel model by adding a third-path along the spatial dimension. A simple but effective attention along the channels is proposed to make TADRN suitable for ad-hoc array processing, \emph{i.e.}, the number of microphones and order invariant processing. Experimental results have established that TADRN is highly effective in utilizing multichannel information even from the microphones at far locations. 


\bibliographystyle{IEEEtran}
\bibliography{mybib}

\begin{thebibliography}{10}
\providecommand{\url}[1]{#1}
\csname url@samestyle\endcsname
\providecommand{\newblock}{\relax}
\providecommand{\bibinfo}[2]{#2}
\providecommand{\BIBentrySTDinterwordspacing}{\spaceskip=0pt\relax}
\providecommand{\BIBentryALTinterwordstretchfactor}{4}
\providecommand{\BIBentryALTinterwordspacing}{\spaceskip=\fontdimen2\font plus
\BIBentryALTinterwordstretchfactor\fontdimen3\font minus
  \fontdimen4\font\relax}
\providecommand{\BIBforeignlanguage}[2]{{%
\expandafter\ifx\csname l@#1\endcsname\relax
\typeout{** WARNING: IEEEtran.bst: No hyphenation pattern has been}%
\typeout{** loaded for the language `#1'. Using the pattern for}%
\typeout{** the default language instead.}%
\else
\language=\csname l@#1\endcsname
\fi
#2}}
\providecommand{\BIBdecl}{\relax}
\BIBdecl

\bibitem{gannot2017consolidated}
S.~Gannot, E.~Vincent, S.~Markovich-Golan, and A.~Ozerov, ``A consolidated
  perspective on multimicrophone speech enhancement and source separation,''
  \emph{IEEE/ACM Transactions on Audio, Speech, and Language Processing}, pp.
  692--730, 2017.

\bibitem{markovichgolan2015optimal}
S.~Markovich-Golan, A.~Bertrand, M.~Moonen, and S.~Gannot, ``Optimal
  distributed minimum-variance beamforming approaches for speech enhancement in
  wireless acoustic sensor networks,'' \emph{Signal Processing}, vol. 107, pp.
  4--20, 2015.

\bibitem{wang2017supervised}
D.~L. Wang and J.~Chen, ``Supervised speech separation based on deep learning:
  An overview,'' \emph{IEEE/ACM Transactions on Audio, Speech, and Language
  Processing}, pp. 1702--1726, 2018.

\bibitem{erdogan2016improved}
H.~Erdogan, J.~R. Hershey, S.~Watanabe, M.~I. Mandel, and J.~Le~Roux,
  ``Improved {MVDR} beamforming using single-channel mask prediction
  networks,'' in \emph{INTERSPEECH}, 2016, pp. 1981--1985.

\bibitem{heymann2016neural}
J.~Heymann, L.~Drude, and R.~Haeb-Umbach, ``Neural network based spectral mask
  estimation for acoustic beamforming,'' in \emph{ICASSP}, 2016, pp. 196--200.

\bibitem{wang2018combining}
Z.-Q. Wang and D.~Wang, ``Combining spectral and spatial features for deep
  learning based blind speaker separation,'' \emph{IEEE/ACM Transactions on
  Audio, Speech, and Language Processing}, pp. 457--468, 2018.

\bibitem{wang2018multi}
Z.-Q. Wang, J.~Le~Roux, and J.~R. Hershey, ``Multi-channel deep clustering:
  Discriminative spectral and spatial embeddings for speaker-independent speech
  separation,'' in \emph{ICASSP}, 2018, pp. 1--5.

\bibitem{wang2020multi}
Z.-Q. Wang and D.~Wang, ``Multi-microphone complex spectral mapping for speech
  dereverberation,'' in \emph{ICASSP}, 2020, pp. 486--490.

\bibitem{tolooshams2020channel}
B.~Tolooshams, R.~Giri, A.~H. Song, U.~Isik, and A.~Krishnaswamy,
  ``Channel-attention dense {U-Net} for multichannel speech enhancement,'' in
  \emph{ICASSP}, 2020, pp. 836--840.

\bibitem{gu2019end}
R.~Gu, J.~Wu, S.-X. Zhang, L.~Chen, Y.~Xu, M.~Yu, D.~Su, Y.~Zou, and D.~Yu,
  ``End-to-end multi-channel speech separation,'' \emph{arXiv:1905.06286},
  2019.

\bibitem{tzirakis2021multi}
P.~Tzirakis, A.~Kumar, and J.~Donley, ``Multi-channel speech enhancement using
  graph neural networks,'' in \emph{ICASSP}, 2021, pp. 3415--3419.

\bibitem{pandey2022tparn}
A.~Pandey, B.~Xu, A.~Kumar, J.~Donley, P.~Calamia, and D.~L. Wang, ``{TPARN:
  T}riple-path attentive recurrent network for time-domain multichannel speech
  enhancement,'' in \emph{ICASSP}, 2022, pp. 6497--6501.

\bibitem{zaheer2017deep}
M.~Zaheer, S.~Kottur, S.~Ravanbakhsh, B.~Poczos, R.~Salakhutdinov, and
  A.~Smola, ``Deep sets,'' \emph{arXiv:1703.06114}, 2017.

\bibitem{liu2008sound}
Z.~Liu, ``Sound source separation with distributed microphone arrays in the
  presence of clocks synchronization errors,'' in \emph{International Workshop
  for Acoustic Echo and Noise Control}, 2008.

\bibitem{araki2018meeting}
S.~Araki, N.~Ono, K.~Kinoshita, and M.~Delcroix, ``Meeting recognition with
  asynchronous distributed microphone array using block-wise refinement of
  mask-based {MVDR} beamformer,'' in \emph{ICASSP}, 2018, pp. 5694--5698.

\bibitem{yoshioka2019meeting}
T.~Yoshioka, Z.~Chen, D.~Dimitriadis, W.~Hinthorn, X.~Huang, A.~Stolcke, and
  M.~Zeng, ``Meeting transcription using virtual microphone arrays,''
  \emph{arXiv:1905.02545}, 2019.

\bibitem{luo2020end}
Y.~Luo, Z.~Chen, N.~Mesgarani, and T.~Yoshioka, ``End-to-end microphone
  permutation and number invariant multi-channel speech separation,'' in
  \emph{ICASSP}, 2020, pp. 6394--6398.

\bibitem{wang2020neural}
D.~Wang, Z.~Chen, and T.~Yoshioka, ``Neural speech separation using spatially
  distributed microphones,'' \emph{arXiv:2004.13670}, 2020.

\bibitem{furnon2021dnn}
N.~Furnon, R.~Serizel, S.~Essid, and I.~Illina, ``{DNN}-based mask estimation
  for distributed speech enhancement in spatially unconstrained microphone
  arrays,'' \emph{IEEE/ACM Transactions on Audio, Speech, and Language
  Processing}, vol.~29, pp. 2310--2323, 2021.

\bibitem{furnon2021attention}
------, ``Attention-based distributed speech enhancement for unconstrained
  microphone arrays with varying number of nodes,'' in \emph{European Signal
  Processing Conference}, 2021, pp. 1095--1099.

\bibitem{zhang2021deep}
X.-L. Zhang, ``Deep ad-hoc beamforming,'' \emph{Computer Speech \& Language},
  vol.~68, p. 101201, 2021.

\bibitem{wang2021continuous}
D.~Wang, T.~Yoshioka, Z.~Chen, X.~Wang, T.~Zhou, and Z.~Meng, ``Continuous
  speech separation with ad hoc microphone arrays,'' \emph{arXiv preprint
  arXiv:2103.02378}, 2021.

\bibitem{bertrand2010distributed}
A.~Bertrand and M.~Moonen, ``Distributed adaptive node-specific signal
  estimation in fully connected sensor networks—{P}art i: Sequential node
  updating,'' \emph{IEEE Transactions on Signal Processing}, vol.~58, no.~10,
  pp. 5277--5291, 2010.

\bibitem{pandey2020dual}
A.~Pandey and D.~Wang, ``Dual-path self-attention {RNN} for real-time speech
  enhancement,'' \emph{arXiv:2010.12713}, 2020.

\bibitem{pandey2022self}
------, ``Self-attending {RNN} for speech enhancement to improve cross-corpus
  generalization,'' \emph{IEEE/ACM Transactions on Audio, Speech, and Language
  Processing}, vol.~30, pp. 1374--1385, 2022.

\bibitem{reddy2020interspeech}
C.~K. Reddy, V.~Gopal, R.~Cutler, E.~Beyrami, R.~Cheng, H.~Dubey,
  S.~Matusevych, R.~Aichner, A.~Aazami, S.~Braun \emph{et~al.}, ``The
  {INTERSPEECH} 2020 deep noise suppression challenge: Datasets, subjective
  testing framework, and challenge results,'' \emph{arXiv:2005.13981}, 2020.

\bibitem{scheibler2018pyroomacoustics}
R.~Scheibler, E.~Bezzam, and I.~Dokmani{\'c}, ``Pyroomacoustics: A python
  package for audio room simulation and array processing algorithms,'' in
  \emph{ICASSP}, 2018, pp. 351--355.

\bibitem{pandey2021dense}
A.~Pandey and D.~Wang, ``Dense {CNN} with self-attention for time-domain speech
  enhancement,'' \emph{IEEE/ACM Transactions on Audio, Speech, and Language
  Processing}, pp. 1270--1279, 2021.

\bibitem{micikevicius2018mixed}
\BIBentryALTinterwordspacing
P.~Micikevicius, S.~Narang, J.~Alben, G.~Diamos, E.~Elsen, D.~Garcia,
  B.~Ginsburg, M.~Houston, O.~Kuchaiev, G.~Venkatesh, and H.~Wu, ``Mixed
  precision training,'' in \emph{ICLR}, 2018. [Online]. Available:
  \url{https://openreview.net/forum?id=r1gs9JgRZ}
\BIBentrySTDinterwordspacing

\bibitem{taal2011algorithm}
C.~H. Taal, R.~C. Hendriks, R.~Heusdens, and J.~Jensen, ``An algorithm for
  intelligibility prediction of time--frequency weighted noisy speech,''
  \emph{IEEE Transactions on Audio, Speech, and Language Processing}, pp.
  2125--2136, 2011.

\bibitem{rix2001perceptual}
A.~W. Rix, J.~G. Beerends, M.~P. Hollier, and A.~P. Hekstra, ``Perceptual
  evaluation of speech quality ({PESQ}) - a new method for speech quality
  assessment of telephone networks and codecs,'' in \emph{ICASSP}, 2001, pp.
  749--752.

\end{thebibliography}


\end{document}